# Charge Effects on Gravitational Wave Detectors

R. F. O'Connell [+]

Department of Physics and Astronomy
Louisiana State University
Baton Rouge, Louisiana  70803-4001

We show that the mean-square displacement of a charged oscillator due to the zero point oscillations of the radiation field is unique in the sense that it is very sensitive to the value of the bare mass of the charge. Thus, a controlled experiment using gravitational wave detectors could lead to a determination of the electron bare mass and shed some light on quantum electrodynamic theory. We also speculate that the irregular signals of non-gravitational origin often observed in gravitational wave bar detectors could be caused by stray charges and that such charges could also adversely affect LIGO and other such detectors.

[+]Electronic address: rfoc@rouge.phys.lsu.edu

1. Introduction

There is intense activity throughout the world directed toward efforts to detect gravitational radiation and we refer to the recent detailed review of Ricci and Brillet [1] for details of same. The initial efforts made use of bar detectors many of which are still functioning. However, it is now recognized that detection schemes based on light interferometry, such as the LIGO effort, offers better prospects of success. In all cases, the goal is to detect displacements due to gravitational radiation.

Bar detectors have been functioning for many years and they often record irregular signals of non-gravitational origin [2]. These signals are relatively large and generally assumed to be non-Gaussian in character. Our purpose here is twofold:

(a) to point out that stray charges (due to ultra-violet radiation, cosmic rays, etc.) impinging on the detector can give rise to such signals and that such charges can adversely affect all gravitational wave detectors. For example, in the case of LIGO, the stray charges would affect the four test masses cum attached mirrors.

(b) to suggest that a controlled experiment using gravitational wave detectors could lead to a determination of the electron bare mass and shed some light on quantum electrodynamic theory.

Our analysis is based on a very general dissipative model which was developed by Ford, Lewis and the present author [3]. In section 2 we present the salient points of our general dissipative model. This model is microscopic in origin and is based on a well-defined Hamiltonian, which in turn leads an equation of motion in the form of a generalized quantum Langevin equation (GLE), in contrast to various specific models found in the literature which adopt at the beginning a phenomenological equation of motion. Next, for purposes of orientation, we consider, in Section 3, the Ohmic heat bath which is universally used to describe bar detectors. This model assumes that $\tilde{\mu}(\omega)$, the Fourier transform of the memory function appearing in the GLE (see (1) below) is frequency independent. In Section 4, we note that thermal effects on the mirror suspension wires of LIGO and other interferometric detectors is generally modeled [1] by a phenomenological function $\phi(\omega)$ and we point out that the well-defined quantity $\tilde{\mu}(\omega)$ is a better quantity to use for this purpose. Finally, in section 5, we analyze charge detectors.

2. General Dissipative Model

In recent years, there has been widespread interest in dissipative problems arising in a variety of areas in physics, an example being gravitational wave detection [1]. As it turns out, solutions of many of these problems are encompassed by a generalization of Langevin's equation to encompass quantum, memory and non-Markovian effects, as well as arbitrary temperature and the presence of an external potential V(x). As in [3], we refer to this as the generalized quantum Langevin equation (GLE):

$$m\ddot{x} + \int_{-\infty}^{t} dt' \, \mu(t-t')\dot{x}(t') + V'(x) = F(t) + f(t) , \qquad (1)$$

where $V'(x) = dV(x)/dx$ is the negative of the time-independent external force and $\mu(t)$ is the so-called memory function. F(t) is the random (fluctuation or noise) force and f(t) is a c-number external force (due to a gravitational wave, for instance). In addition (keeping in mind that measurements of x generally involve a variety of readout systems involving electrical measurements), it should be strongly emphasized that "-- the description is more general than the language --" [3] in that x(t) can be a generalized displacement operator (so that, for instance, x could represent a voltage change).

A detailed discussion of Eq. (1) appears in [3]. In particular, it was pointed out the GLE corresponds to a description of a quantum system interacting with a quantum-mechanical heat bath and that this description can be precisely formulated, using such general principles as causality and the second law of thermodynamics. We also stressed that this is a model-independent description. However, the most general GLE can be realized with a simple and convenient model, viz. the independent-oscillator (IO) model.

The Hamiltonian of the IO system is

$$H = \frac{p^2}{2m} + V(x) + \sum_j \left[ \frac{p_j^2}{2m_j} + \frac{1}{2} m_j \omega_j^2 (q_j - x)^2 \right] - xf(t) . \tag{2}$$

Here m is the mass of the quantum particle while $m_j$ and $\omega_j$ refer to the mass and frequency of heat-bath oscillator j. In addition, x and p are the coordinate and momentum operators for the quantum particle and $q_j$ and $p_j$ are the corresponding quantities for the heat-bath oscillators.

The infinity of choices for the $m_j$ and $\omega_j$ give this model its great generality. In particular, it can describe non-relativistic quantum electrodynamics, the Schwabl-Thirring model, the Ford-Kac-Mazur model and the Lamb model [3]. Use of the Heisenberg equations of motion lead to the GLE, Eq. (1), describing the time development of the particle motion, with

$$\mu(t) = \sum_j m_j \omega_j^2 \cos(\omega_j t) \theta(t) , \tag{3}$$

where (t) is the Heaviside step function. Also

$$F(t) = \sum_j m_j \omega_j^2 q_j^h(t), \tag{4}$$

where $q^h(t)$ denotes the general solution of the homogeneous equation for the heat-bath oscillators (corresponding to no interaction). This solution of (1) is readily obtained when $V(x) = 0$, corresponding to the original Brownian motion problem [4]. As shown by FLO [5] a solution is also possible in the case of an oscillator. Taking,

$V(x) = \frac{1}{2} m \omega_0^2 x^2$, these authors obtained (see equations (1) to (3) of Ref. 5)

$$\tilde{x}(\omega) = \alpha(\omega)\{\tilde{F}(\omega) + \tilde{f}(\omega)\} , \tag{5}$$

where

$$\alpha(\omega) = [-m\omega^2 + m\omega_0^2 - i\omega\tilde{\mu}(\omega)]^{-1} , \tag{6}$$

is the generalized susceptibility and the superposed tilde is used to denote the Fourier transform. Thus, $\tilde{x}(\omega)$ is the Fourier transform of the operator x(t):

$$\tilde{x}(\omega) = \int_{-\infty}^{\infty} dt\, x(t) e^{j\omega t} . \tag{7}$$

Also, since Eq. (3) implies that μ(t) is 0 for negative t, the Fourier transfer of the memory function is given by

$$\tilde{\mu}(\omega) = \int_0^{\infty} dt\, \mu(t) e^{i\omega t}, \; \text{Im}\,\omega > 0 . \tag{8}$$

Thus $\tilde{\mu}(\omega)$ is analytic in the upper half-plane, $\text{Im} > 0$.

We have now all the tools we need to calculate various correlation functions which represent, in essence, observable quantities. In particular, we obtain the coordinate autocorrelation function

$$C(t) = \frac{1}{2}\langle x(t)x(0) + x(0)x(t)\rangle = \frac{\hbar}{\pi} \int_0^{\infty} d\omega\, \text{Im}\{\alpha(\omega + i0^+)\} \coth\frac{\hbar\omega}{2kT}\cos\omega t , \tag{9}$$

where $\alpha(\omega)$ is given by (6). Our focus is on $\langle x^2 \rangle$, the mean-square displacement of the oscillator due to the heat bath, which may be written in the form

$$<x^2> = C(0) = \int_0^{\infty} d\omega\, P(\omega) , \tag{10}$$

where, from (9) and (6), we see that P( ), the power spectrum of the coordinate fluctuations, may be written in the form

$$P(\omega) = \frac{\hbar}{\pi} \operatorname{Im}\alpha(\omega) \coth\left(\frac{\hbar\omega}{2kT}\right)$$

$$= \frac{\hbar}{\pi} \frac{\omega \operatorname{Re}\tilde{\mu}(\omega) \coth(\hbar\omega/2kT)}{m^2\left\{\omega^2 - \omega_0^2 - \frac{\omega}{m}\operatorname{Im}\tilde{\mu}(\omega)\right\}^2 + (\omega \operatorname{Re}\tilde{\mu}(\omega))^2} \quad . \tag{11}$$

Now, as discussed in [3], $\tilde{\mu}(z)$ is not only analytic in the upper half-plane, Im z > 0, but it is a <u>positive real function</u> with the consequence that the relation between its real and imaginary parts is given by the Stieltjes inversion theorem i.e. a Kramers-Kronig relation with at most one subtraction term, which can be absorbed into the particle mass term. Thus, in essence, $\operatorname{Re}\tilde{\mu}(\omega)$ characterizes the function $\tilde{\mu}(\omega)$ and hence it is the key ingredient in determining observable results. Furthermore, as derived explicitly in [3], we have

$$\operatorname{Re}\left[\tilde{\mu}(\omega + i0^+)\right] = \frac{\pi}{2}\sum_j m_j \omega_j^2 \left[\delta(\omega - \omega_j) + \delta(\omega + \omega_j)\right] , \tag{12}$$

and thus we see explicitly that resonances also occur at the normal-mode frequencies of the heat-bath in addition to the resonance at $\omega \approx \omega_0$ (assuming $\operatorname{Im}\tilde{\mu}(\omega) << \operatorname{Re}\tilde{\mu}(\omega)$, which is generally the case). This completes our discussion of our general dissipative model. It has been applied to a wide variety of problems in many areas of physics (see [4] for a review) but here, of course, our emphasis is on gravitational wave detectors. For orientation purposes, we first consider the "Ohmic" heat bath (for which $\operatorname{Im}\tilde{\mu}(\omega) = 0$) which is universally used to describe bar detectors and then go on to consider interferometric detectors and, finally, charged detectors.

3. Ohmic Heat Bath

The detector (bar, mirror, etc.) is conventionally described as an oscillator of mass m and natural frequency $\omega_0$ which is immersed in a dissipative environment (heat bath) at temperature T. The bar detector is a resonant detector sensitive only to a narrow band of frequencies centered at $\omega \approx \omega_0$. Hence, it is permissible to take $\tilde{\mu}(\omega) = m\gamma$, a constant. Then, at T = 0, from (6), (10) and (11), we obtain

$$\langle x^2 \rangle = \frac{\hbar\gamma}{\pi m} \int_0^\infty d\omega \frac{\omega}{(\omega^2 - \omega_0^2)^2 + \gamma^2 \omega^2} = \frac{\hbar}{\pi m \omega_1} \sin^{-1}(\omega_1/\omega_0) , \qquad (13)$$

where $\omega_1 = \{\omega_0^2 - (\gamma^2/4)\}^{1/2}$. For weak coupling $(\gamma << \omega_0)$, which is the case for gravitational detectors,

$$\langle x^2 \rangle \approx (\hbar/2 m\omega_0) \equiv \langle x^2 \rangle_Q , \qquad (14)$$

the familiar quantum mechanical result for the uncertainty in the position of a free ($\gamma = 0$) oscillator at T = 0, and which leads to a "standard quantum limit" for the detection of gravitational radiation [6]. We see from (13) that $\langle x^2 \rangle$ monotonically decreases with increasing $\gamma$ but for gravitational detectors this is inconsequential. In general, higher temperatures lead to larger values of $\langle x^2 \rangle$ and, in the limit $\hbar\omega_0 << kT$, (9) leads to the familiar result

$$\langle x^2 \rangle \approx (kT/m\omega_0^2) = (2kT/\hbar\omega_0)\langle x^2 \rangle_Q . \qquad (15)$$

Since, in general, $\gamma << \omega_0$, it is clear from (13) that the spectrum is highly peaked at the resonant frequency $\omega \approx \omega_0$ which is why bar detectors are essentially resonant detectors.

4. Heat Bath for Interferometric Mirrors

Non-resonant interferometric detectors such as LIGO, are responsive to a range of frequencies, and thus the frequency dependence of $\tilde{\mu}(\omega)$ is essential. The existing analysis makes use of the phenomenological Zener function $\phi(\omega)$, as discussed in [1].

Thus, in order to make contact with the approaches which use the Zener function, we write $k = m\omega_0^2$, so that the generalized susceptibility (6) may be written as

$$\alpha(\omega) = \{-m\omega^2 + k_{eff}\}^{-1} \ , \tag{16}$$

where

$$k_{eff} = k\left\{1 - \frac{i}{k}\omega\tilde{\mu}(\omega)\right\} \equiv k\{1 - i\phi(\omega)\} \ . \tag{17}$$

In other words, the quantity $\phi(\omega)$ appearing in phenomenological theories is simply given by

$$\phi(\omega) = \frac{1}{k}\omega\tilde{\mu}(\omega) \ . \tag{18}$$

However, from our perspective, it is not helpful to discuss the results in terms of an effective complex spring constant, $k_{eff}$, since from (3) we see that $\mu(t)$ depends only on the parameters of the heat bath. In fact, as emphasized in [3], the memory function is independent of the external potential, the particle mass and the temperature T. In other words, the imaginary part of $k_{eff}$ does not depend in any way on the properties of the spring (apart from an overall non-essential factor of m) but, instead, it depends on the nature of the dissipative environment (primarily the suspension wire in the case of LIGO).

There are several key reasons why it is better to fit the experimental results by using $\mu(\omega)$ instead of $\phi(\omega)$ (apart from the fact that (17) could be misleading since it tends to obscure the fact that $\phi(\omega)$ itself can also have an imaginary part). These stem from the fact that we know a lot about the properties of $\tilde{\mu}(\omega)$, regardless of the nature of the heat bath. In particular, as discussed, after (11) in section 2, it is a positive real function. In addition, as we saw in (12), $\text{Re}\,\tilde{\mu}(\omega)$ is given explicitly in terms of the parameters of the heat-bath. Thus, in the case of LIGO, this will give information on the nature of the dissipative effect of the suspension wires. All of these properties should be a guide to the experimentalist in choosing suitable parameters to fit the data.

**5.** Radiation heat bath

If the oscillator-detector acquires a stray charge, q say, then it will interact with the electromagnetic fields associated with the ambient blackbody radiation. In particular, $\tilde{\mu}(\omega)$ has a strong frequency dependence. The analysis in this case is non-trivial and involves the usual mass renormalization of QED [3, 5, 7].

In fact, as shown explicitly in [3], the Hamiltonian (2) of our general dissipative model incorporates non-relativistic QED as a particular case. In addition, our analysis incorporated a form factor $f_k$ for the charge (where $f_k$ is the Fourier transform of the charge distribution) in order to overcome the well-known problems (such as runaway solutions and lack of causality (7)) associated with the motion of a point charge. We used a form factor similar to that used by other investigators [8, 9] but the final results for the equations of motion of a radiating electron were not sensitive to the choice [7]. Explicitly, we have [3, 5, 7]

$$\tilde{\mu}(\omega) = \frac{M\tau_q \Omega^2}{\omega^2 + \Omega^2} \omega(\omega - i\Omega) . \tag{19}$$

where M is the renormalized (physical) mass of the charge q, $\Omega$ is a large cut-off frequency and $\tau_q = (2q^2 / 3Mc^3)$ which is equal to $6 \times 10^{-24}$ s for the electron charge. Causality demands that $\Omega \leq \tau_q^{-1}$ and the choice $\Omega = \tau_q^{-1}$ ($1.2 \times 10^{23}$ s$^{-1}$ for the electron charge) corresponds to a charge with the smallest possible size consistent with causality [7]. This results in m = 0 where m is the bare mass of the charge. This is the usual choice for the bare mass of the electron in QED [3, 7] but, more generally, we have the explicit relation for the relation between M, m and $\Omega$ [7],

$$M = m + \tau_q \Omega M . \tag{20}$$

Thus, we see immediately that for $\Omega = \tau_q^{-1}$ we have m = 0 but a choice of $\Omega$ less than $\tau_q^{-1}$ would lead to a positive non-zero value for m. As we will now show, the result for the mean-square displacement of a charged oscillator due to interaction with the zero-point oscillations of the radiation field is unique in the sense that it is very sensitive to the value of m (in contrast to the case of the equation of motion of the radiating charge which is not sensitive to the value of m). An exact result for $<x^2>$ may be obtained for all temperatures, by substituting (19) in (11) and using (10), but, for our purposes, it is sufficient to display the dominant term in the T = 0 result (since finite temperature effects should be even larger, as we saw explicitly in section 3) viz. [10]

$$\langle x^2 \rangle = \frac{\hbar \tau_q}{\pi M} \log(M / m_0 \tau_q)$$

$$= \frac{2}{\pi}(\omega_0 \tau_q) <x^2>_Q \log(M / m\omega_0 \tau_q) \ . \tag{21}$$

In the second equality we have written the result in terms of $<x^2>_Q$, the aforementioned result for an uncharged quantum oscillator of mass M, given in (14).

In the case of the LSU bar detector, for example, we have M = 1,184 kgs and $\omega_0 = 5.7 \times 10^3 s^{-1}$ so that $\{<x^2>_Q\}^{1/2} = 3 \times 10^{-19}$ cm. It also follows that, if we take T = 4.2 K, the rms displacement corresponding to (15) is larger by a factor of $3.2 \times 10^4$. In addition, if we take the capacitance of the bar (whose length is 300 cm) to be, say, 100 pf and let it charge to 10V then $q = 10^9$ C $= 6.25 \times 10^9 |e|$, where e is the electronic charge, and hence $\tau_q = 2 \times 10^{-37}$ s so that $\omega_0 \tau_q = 10^{-33}$. [It should be noted that the small quantity $\omega_0 \tau_q$ also appears inside the log factor in the denominator]. Hence, the rms value of the factor multiplying the log term in (21) is $8 \times 10^{-36}$ cm, a very small number. More significant is the fact that m appears in the denominator of the log factor. Since $-\log m \to \infty$ as $m \to 0$, the rms displacement $\sqrt{<x^2>}$ is actually logarithmatically divergent for m = 0. We speculate that this divergence could be tempered by a non-zero m (which would have implications for mass renormalization theory in QED) or by the presence of anharmonic terms which will come into play as $\langle x^2 \rangle$ gets large. It may also point to the necessity of extending the theoretical framework to incorporate retardation and relativistic effects. In addition, we stress that the underlying theory is the same as we used in the derivation of an equation of motion of a radiating electron [7] and which, in contrast to the Abraham-Lorentz equation, is second-order, has no runaway solutions and obeys the fluctuation-dissipation theorem. Further discussion of our results for the equation of the radiating electron may be found in the new edition of Jackson's book [11]. However, with respect to the present calculation, while we feel that the derivation presented is the best presently available, it is clear to us that further work is warranted. In essence, we are in the realm of macroscopic quantum mechanics, a relativity new area of inquiry [12]. In addition, while there should be some differences in the behavior of metals (such as the Al bar detectors) and insulators (such as the proposed fused silica or sapphire LIGO masses), because of the difference in charge distribution, we expect that such differences will be small compared to the dominant effect considered here. On the theoretical side, it might prove profitable to make use of the relativistic equation for the radiating electron [13] and also take retardation effects into account.

In summary, we have demonstrated a possible origin for the relatively large ubiquituous non-gravitational signals observed by bar detectors and raised a flag indicating that such effects may also be present in LIGO and other detectors. Moreover, with regard to the LIGO experiment, we pointed out the advantages of using $\tilde{\mu}(\omega)$ to parameterize the experimental results instead of the currently used Zener function $\phi(\omega)$ [1, 14, 15]. In addition, we have identified a possible test for determining m, the bare mass of the electron. Thus, it would be of interest to investigate such possibilities experimentally by manually charging the detectors in a controlled experiment.

Finally, we note a recent paper by Astone et al. [16] in which they observed mechanical vibrations in the resonant gravitational wave detector NAUTILUS due to the passage of cosmic rays; these authors attribute the effect to warming up of the bar but perhaps charge effects are playing a role.

**Acknowledgment**

The author is grateful to W. Johnson for an extended and enlightening discussion based on his own experience with gravitational bar detectors and his knowledge of the work of other experimental groups. He also thanks P. Adams for a useful conversation. He would especially like to thank Professor G. W. Ford for continued discussions on the fundamentals underlying the subject.

References

1. F. Ricci and A. Brillet, Annu. Rev. Nucl. Part. Sci. **47** (1997) 111.

2. W. Johnson, private communication.

3. G. W. Ford, J. T. Lewis, and R. F. O'Connell, Phys. Rev. A, **37**, 4419 (1988).

4. R. F. O'Connell, "Dissipative and Fluctuation Phenomena in Quantum Mechanics with Applications", in Festschrift for J. P. Dahl, Int. J. Quantum Chem., **58**, 569 (1996).

5. W. Ford, J. T. Lewis, and R. F. O'Connell, Phys. Rev. Lett., **55**, 2273 (1985).

6. V. B. Braginsky and Yu. I. Vorontsov, Usp. Fiz. Nauk. **114** (1974) 41 [Sov. Phys. Usp. 17 (1975) 644].

7. G. W. Ford and R. F. O'Connell, Phys. Lett. A **157** (1991) 217; ibid. Phys. Rev. A **57** (1998) 3112.

8. C. Cohen-Tannoudji, J. Dupont-Roc and G. Grynberg "Introduction to Quantum Electrodynamics" (J. Wiley and Sons, New York, 1989), p. 38.

9. P. de Vries, D. V. van Coevorden and A. Lagendijk, Rev. Mod. Phys. **70**, 447 (1998), eq. (24).

10. X. L. Li, G. W. Ford and R. F. O'Connell, Physica A **193** (1993) 575. See especially Eqs. (2.4) and (3.13).

11. J. D. Jackson, <u>Classical Electrodynamics</u>, 3$^{rd}$ ed. (Wiley, New York, 1998), pps. 749 and 772.

12. A. O. Caldeira and A. J. Leggett, Ann. Phys. (NY) **149** (1983) 374.

13. G. W. Ford and R. F. O'Connell, Phys. Lett. A **174** (1993) 182.

14. P. R. Saulson, Phys. Rev. D **42** (1990) 2437.

15. G. I. González and P. R. Saulson, J. Acoust. Soc. Am., **96** (1994) 207; ibid. Phys. Lett. A, **201** (1995) 12.

16. P. Astone et al., Phys. Rev. Lett. **84** (2000) 14.